\documentclass[journal]{IEEEtran} %
\usepackage[utf8]{inputenc}
\usepackage[T1]{fontenc}
\usepackage{microtype}
\usepackage[hidelinks]{hyperref}
\usepackage{url}
\usepackage{graphicx}
\usepackage[breakable,skins]{tcolorbox}
\usepackage{balance}

\begin{document}

\title{Agile beyond teams and feedback beyond software in automotive systems} %
\author{S. Magnus Ågren, Rogardt Heldal, Eric Knauss, and Patrizio Pelliccione
\thanks{S. M. Ågren is with the Department of Computer Science and Engineering at Chalmers $\mid$ University of Gothenburg, Sweden (e-mail: magnus.agren@chalmers.se).}
\thanks{R. Heldal is with the Western Norway University of Applied Sciences and the Department of Computer Science and Engineering at Chalmers $\mid$ University of Gothenburg, Sweden (e-mail: rogardt.heldal@hvl.no).}
\thanks{E. Knauss is with the Department of Computer Science and Engineering at Chalmers $\mid$ University of Gothenburg, Sweden (e-mail: eric.knauss@gu.se).}
\thanks{P. Pelliccione is with the Gran Sasso Science Institute (GSSI), Italy, and the Department of Computer Science and Engineering at Chalmers $\mid$ University of Gothenburg, Sweden (e-mail: patrizio.pelliccione@gssi.it).}
\thanks{We thank all respondents in the study for their valuable input.
We also thank Gösta Malmqvist, Jonas Bodén, Andreas Karlsson, and Caroline Svensson at Knowit AB, for their help with setting up and performing the interviews.
This study was performed in collaboration with the Vinnova project Next Generation Electrical Architecture (NGEA).
This work was also partially supported by the Software Center project \emph{Engineering Knowledge-Flows in Large-Scale Agile Systems Development}.}%
}
\date{September 2020}

\maketitle

\begin{abstract} %

In order to increase the ability to build complex, software-intensive systems, as well as to decrease time-to-market for new functionality, automotive companies aim to scale agile methods beyond individual teams.
This is challenging, given the specifics of automotive systems that are often safety-critical and consist of software, hardware, and mechanical components.
This paper investigates the concrete reasons for scaling agility beyond teams, the strategies that support such scaling, and foreseeable implications that such a drastic organizational change will entail.
The investigation is based on a qualitative case study, with data from 20 semi-structured interviews with managers and technical experts at two automotive companies.
At the core of our findings are observations about establishing an agile vehicle-level feedback loop beyond individual teams.  
(I) We find that automotive OEMs aim to decrease lead-time of development.
(II) We also identify 7 strategies that aim to enable scaled-agile beyond teams.
(III) Finally, we extract 6 foreseeable implications and side-effects of scaling agile beyond teams in automotive.
By charting the landscape of expected benefits, strategies, and implications of scaling agile beyond teams in automotive, we enable further research and process improvements.
\\

\emph{Managerial Relevance Statement}---From interviews with high- and mid-level managers in two automotive OEMs, we derive qualitative insights on scaling agile ways-of-working beyond single teams.
We analyze what abilities automotive OEMs are seeking to achieve from applying agile at this scope, strategies proposed for scaling, and implications anticipated of the strategies.
We believe that the findings of this study can provide guidance, on the trade-offs between different approaches when scaling agile, to managers in automotive or other domains with a mix of software, hardware, and mechatronics.
\end{abstract}

\section{Introduction}
Agile ways-of-working initially focused on small teams developing software.
The success of agile approaches in this context has, however, led to adoption also in the development of large-scale~\cite{uludag2020revealing} and mechatronic systems, where a stage-gate process has been the norm~\cite{pernstal12}.
One example is the automotive domain.
New technologies, such as electrification, autonomous driving, and connectivity are increasing the focus on software at automotive Original Equipment Manufacturers (OEMs).
This is also testified by leading software companies, like Google and Apple, entering this market.
To remain competitive, OEMs seek to reduce time-to-market and increase flexibility -- the ability to rapidly react to change -- through increased development speed.
Agile ways-of-working do that by fast and early feedback on the product level.
However, compared with software, development of hardware and mechanical components has long lead-times.
On the whole-system level, the pace of integration and, consequently, the length of the feedback loop are determined by the longest lead-time~\cite{berger15}.
Under such conditions, scaling agile to multiple teams involves several challenges~\cite{eklund2014industrial}: (i) handling system quality requirements, e.g. safety;
(ii) the inherently long feedback loop from customers when financial transactions are tied to delivery of physical products;
(iii) mechanical development striving for long-term predictability, whereas software development strives for short-term agility, and (iv) handling the technical complexity, and associated expertise needed, of interdependent components in large-scale systems.

This paper reports on a case study in the automotive domain.
At our case companies, work is largely done in teams working according to an agile method, but their surrounding context is specification-driven development, with a stage-gate process.
Related to challenges (ii) and (iv) above, a stage-gate development process has an impact on the applicability of agile ways-of-working when vehicle functions span the work of multiple teams.
For example, developing an autonomous driving function could involve teams working on the engine, the braking system, sensors such as cameras and radars, as well as teams working on the driving algorithms.
Assuming a stage-gate function development process, the feedback loop for each team within agile sprints does not go all the way to full system integration.
Such a full-system feedback loop requires scaling agile beyond individual teams, and this implies scaling agile also beyond software.

Although the agile movement originated from developers -- as expressed for example in the opening words of the agile manifesto: \emph{``We are uncovering better ways of developing
software by doing it''}~\cite{agilemanifesto} -- when agile ways-of-working are to cover large parts of an entire organization, this impacts also the manager roles.
They are immediately affected by, and sometimes also responsible for leading, these changes.
As a complement to a developer-centric view, we seek the view of managers, whose roles involve multiple teams.
Striving to adopt agility at the company level in automotive requires addressing non-agile surroundings of individual teams, although the teams themselves may be internally agile.
With organizational aspects affected, agility spanning multiple teams goes beyond technical aspects, and thus beyond software.

A challenge when reasoning about agile methods (see Section~\ref{sec:auto-agile}) is the absence of a strict operationalized definition of agility.
In this work, we investigate a knowledge gap concerning scaling agility beyond single teams.
Specifically, we investigate what companies are seeking to achieve, how they envision doing it, and what implications they foresee from such drastic organizational change.
Our study contributes to the understanding of these aspects of agility, in the specific context of the automotive domain.
In particular, we focus on the aspect of fast and early feedback on the product level during development.
Agile feedback loops are at the core of agility since they allow agile teams to react to change~\cite{gren2020agility}.
Therefore, when scaling beyond individual teams and software, we believe that a system-level feedback loop must be at the core of our investigation.
As Vöst~\cite{vost2015vehicle} describes for the case of continuous integration in automotive, the system-level is usually the level of the entire vehicle.
Agile development could either relate to a specific vehicle model or to a repository of components relevant to many vehicle models.
Either way, the feedback loop on any agile development task should relate to properties that are observable on the entire system, that is, the vehicle-level.
We investigate this in detail by conducting a case study at two automotive OEMs.
To elicit respondents' views, we pose the following research questions.

\begin{description}
\item[RQ1] What abilities are automotive OEMs seeking to achieve through an agile vehicle-level feedback loop beyond individual teams?
\item[RQ2] How are automotive OEMs proposing that an agile vehicle-level feedback loop beyond individual teams can be established?
\item[RQ3] What implications do automotive OEMs foresee from the proposed ways of introducing an agile vehicle-level feedback loop beyond individual teams? %
\end{description}

Given our method of thematic analysis of qualitative interview data, our results consist of a number of themes for each research question, Figure~\ref{fig:themes} gives an overview.

\subsubsection*{Paper outline}
The rest of this paper is structured as follows.
Section~\ref{sec:bg} elaborates on software development and agility in our context of the automotive domain.
Section~\ref{sec:method} describes our research method, including the case in our context.
Section~\ref{sec:results} presents our results and Section~\ref{sec:interpretation} discusses how we interpret them.
Section~\ref{sec:future} outlines directions for future research.
Finally, Section~\ref{sec:conclusion} concludes the paper.

\section{Theoretical Background}\label{sec:bg}

The theoretical background covers three aspects, the increase of software in the automotive industry, the impact on architecture, and attempts to handle this change via agile development.
Automotive system offer clear examples of products that require change because of software increases, but where, for example, regulatory and safety concerns add complexity to agile ways-of-working.

\subsection{Software in Automotive Systems}
From a software engineering perspective, the automotive domain combines several traits that impact software development.
The systems are mechatronic, but the amount of software is increasing at an accelerating rate~\cite{hiller_icsa_keynote,ebert2009embedded}.
A modern vehicle can contain more than 100 Electronic Control Units (ECUs) and more than 100M lines of code.
The products are sold on the consumer market, but are also safety-critical.
Taken together, these traits make the automotive domain an interesting context for researching software engineering, also the aspects going beyond the software itself, such as feedback to agile teams from the development of the entire vehicle.

\subsection{Electrical and Software Architecture of Automotive Systems}
In order to describe the specificity of software development in the automotive domain, we describe the evolution of electrical and software architectures of automotive systems. In the automotive domain, software architectures are described at three different levels of abstraction: functional, logical, and technical~\cite{TUM-I0915}. A functional architecture provides a black-box description of the functional structure of the system. A logical architecture is still hardware agnostic and it describes a decomposition of the system into logical components realizing the functionalities formalized through the functional architecture. A technical architecture specifies how the logical components are implemented and then integrated into a hardware platform.
The work in~\cite{BucaioniICSA2020} presents a historical perspective of technical reference architectures and specifically introduces three generations\footnote{Indeed the division into three generations is a simplification and there could exist architectures that are intermediate between two different generations.}~\cite{Bosch2019,Bosch2016}.

\subsubsection{Distributed E/E Architecture -- the traditional architecture}
When electronics started having an important role in vehicles, specific functionalities had been implemented and deployed on specific ECUs. This created a strong coupling between software and hardware since the focus was on developing a specific functionality that was then released as a physical ECU enhanced with the required software. The growing importance of electronics and software led to a growing number of ECUs organized in a distributed and high-modular architecture, where each function is delivered using a specific ECU. 
In this architecture, the integration among different ECUs is obtained by wiring, however, typically, there is limited to no interaction between the different ECUs. Examples of well-known distributed E/E architectures are the Volkswagen Group B platform~\cite{volkb}, the Volvo P80 platform~\cite{volvo}, and the Ford EUCD platform~\cite{ford}.

\subsubsection{Domain Centralised E/E Architecture -- the ``in development" architecture}
The increasing number of ECUs and the attention to software qualities as scalability, robustness, and maintainability led to a new and more structured architecture~\cite{BucaioniICSA2020}. The Domain Centralised E/E architecture adopts a layered architectural style and introduces the concept of domain~\cite{DomainArchitecture}, which is used for grouping ECUs, e.g. Power-train for the control of engine and batteries, Infotainment for the control of displays, entertainment system, and navigation, etc. The Domain Centralised E/E architecture represents the present of automotive architectures. Examples of well-known domain centralized E/E architectures are the Volkswagen Group MQB platform~\cite{volkMQB}, the Volvo Scalable Product Architecture (SPA) platform~\cite{spa}, and the Volkswagen Group MEB platform~\cite{volkMEB}, which has been adopted also by Ford~\cite{fordvv}.  The worldwide standardized software framework of Autosar\footnote{Autosar -- \url{https://www.autosar.org}} 
has been introduced with the aim of triggering a paradigm shift from an ECU-based to a function-based system design. Autosar introduced also an ECU abstraction, i.e. providing a software interface to the electrical values of any specific ECU, in order to decouple software from hardware~\cite{autosar}. However, the decoupling between software and hardware is getting attention but is not completely achieved.

\subsubsection{Vehicle Centralised E/E Architecture -- the ``near future" architecture}\label{sec:future_arch}
This architecture  represents the future of vehicle architectures. One of the main elements of this architecture is the High-Performance Computing (HPC) server, which will be the actual brain of future vehicles. This architecture follows a layered architectural style as the domain centralized E/E architecture.
This architecture exploits technologies like artificial intelligence, over-the-air (OTA) updates, cloud, etc. Examples of vehicle centralized E/E architectures are BMW~\cite{BMWarch3} and Volvo cars (SPA2)~\cite{hiller_icsa_keynote}. The decoupling between software and hardware is a must in this architecture since the most challenging and innovative functionalities will be deployed to one HPC server, which will be mostly under the direct control of the OEM and will make feasible also software upgrade and OTA updates.

\subsection{Agility in Automotive Software Development}\label{sec:auto-agile}
When reasoning about agile methods, a major difficulty is the lack of a working definition on what agile methods are (and particularly what methods are not to be characterized as agile) \cite{helena,Meyer2014}.
Often, the goal of transitioning to agile relates to the hope to increase development speed, i.e. reducing the time from accepting a change request until the changed system is available to users. 
When transitioning to agile ways-of-working, various frameworks have been suggested, including for example the Stairway to Heaven \cite{olsson2012climbing}.
This framework suggests that in order to fully leverage the benefits of agile methods, systems companies need to move from agile teams in their R\&D departments to system level continuous integration, and then on to continuous deployment and organization wide innovation systems that fully leverage the potential of software-based innovations \cite{olsson2012climbing}.
With each step on this stairway, more parts of an organization need to be involved, and in particular, agility must be supported beyond the team.
This has been found challenging \cite{Kasauli2020b}, especially in the automotive sector \cite{agileislands}.
In particular, the goal to provide engineers with continuous feedback on system level becomes quickly difficult and needs extensive investments in infrastructure and tools \cite{knauss2016continuous}.

\section{Method}\label{sec:method}
Research method selection inherently contains a tradeoff between generalizability over actors, precision in measuring behaviors, and realism of context~\cite{stol2018abc}.
As our research questions concern the real-world context of automotive OEMs, we strive to maximize realism of context.
In the taxonomy of Stol and Fitzgerald, research methods that offer maximum potential for realism of context are categorized as field studies~\cite[p11--12]{stol2018abc}.
From this category, the particular method we use is the exploratory case study.
Case studies allow exploring \emph{what's going on} and \emph{how things work} \cite{stol2018abc}, and are suitable when the boundary between phenomenon and context may be unclear~\cite{runeson_case_study,yin2003case}.
For specific case study guidelines, we use Runeson and Höst~\cite{runeson_case_study}, and below we describe our case, data collection and analysis, and discuss the validity and limitations of our study.

\subsection{Case description}
Our case consists of two companies, both being automotive OEMs, one car manufacturer and one heavy vehicles manufacturer.
Both companies are based in Sweden, but have large, global, organizations.
They produce several different models, and they have had a long history of many different owners.
Most projects have a budget spanning from four to nine digits in USD.
Project scopes vary, from minor adjustments of a product for a specific market to entire new vehicles.
A project delivery typically consists of developments in mechanics, hardware, and software technologies.
The release processes are set up to serve major market introductions every few years.
The company cultures are finance- and commitment-oriented, with a strong focus on a phase-gate process.
Embedded systems play a key role, however, software itself has become increasingly important.
We selected these companies as our case since both are actively seeking to establish agile ways-of-working encompassing more of the organization than single, separate, teams.
Teams in our context typically consist of 5--10 persons.
To this end, the Scaled Agile Framework (SAFe)~\cite{leffingwell2016safe} is being adopted in their development organizations.
Our research aims to make explicit which considerations apply when pursuing an agile vehicle-level feedback loop beyond individual teams, through the SAFe framework (or similar) in this context and domain.

We have focused our selection of respondents on high- and mid-level managers since they are, at the organization levels, spanning multiple teams.
At both case companies, we targeted comparable software development intensive system areas, selected for developing central parts of the vehicle electronic system, hence being at the core of the vehicle software, rather than developing services on top of the vehicle.
We interviewed the managers of these areas and managers of all immediate subordinate departments.
The sampling strategy was thus to exhaustively cover corresponding parts of both company management structures.
The exact organization structure differs between the companies, hence there are more respondents from case company one.
The managers' view was complemented by interviews with technical experts from the development organizations of each company.
Table \ref{tab:respondents} gives an overview of our respondents, their roles and years of experience.
In total 20 respondents were interviewed, over the two case companies, with each interview lasting approximately one hour.
For confidentiality, respondents are kept anonymous and referred to with running ids R1 -- R20.

\begin{table}[!t]%
    \centering
    \caption{Selection of Respondents}
    \label{tab:respondents}
    \begin{tabular}{l|l|l}
        & Role & Automotive experience \\
        \hline
        \multicolumn{3}{c}{Case company one} \\
        \hline
        R1 & Technical expert, architecture & $>$30 years \\
        R2 & Manager SW dept. & $>$25 years \\
        R3 & Manager SW dept. & 23 years \\
        R4 & Manager SW dept. & N/A \\
        R5 & Manager mechanical dept. & $>$20 years \\
        R6 & Manager SW dept. & 15 years \\
        R7 & Technical expert, architecture & 18 years \\
        R11 & Manager SW dept. & 1 year \\
        R15 & Manager mechanical dept. & $>$20 years \\
        R16 & Technical expert, process & $>$10 years \\
        R17 & Manager SW dept. & $>$5 years \\
        R18 & Manager SW dept. & $>$20 years \\
        R19 & Manager SW group & N/A \\
        \hline
        \multicolumn{3}{c}{Case company two} \\
        \hline
        R8 & Technical expert, process & 20 years \\
        R9 & Manager SW dept. & 10 years \\
        R10 & Manager SW group & $>$5 years \\
        R12 & Manager SW dept. & $>$25 years \\
        R13 & Manager SW dept. & 21 years \\
        R14 & Manager SW tool dept. & 12 years \\
        R20 & Manager system dept. & 19 years \\
    \end{tabular}
    \vspace{-.5cm}
\end{table}

\subsection{Data collection and analysis}
For data collection, we employed a qualitative approach, using semi-structured interviews.
Semi-structured interviews employ an interview guide with questions, but allow the order of questions to vary to fit the natural flow of the conversation.
All interviews were conducted by the first author, together with one or more co-interviewers from a company active as both a supplier to the two case companies and as an industrial partner company in the research project in which this study was conducted.
The interviews took place in 2017, from February to July.
All interviews were recorded and transcribed.
Before each interview, we acquired consent from the respondent to use their responses in anonymized form.

Our interview questions\footnote{The interview guide used is available online at \url{https://doi.org/10.5281/zenodo.1299206}.} took an exploratory approach.
We asked respondents about their current situation and asked broadly about notions related to agile development, such as team autonomy, organizational and technical dependencies, and development speed.
From our discussions, we have then refined our guiding research questions: RQ1 is about lead-time, rather than development speed, to improve accuracy. 
Enabling strategies (RQ2) and foreseeable implications (RQ3) are directly related to our questions on organizational and technical dependencies in relation to agility.
Parts of the data related to requirements engineering (approximately 15\% of the entire data set) have been used in previous publications~\cite{aagren2018manager,aagren2019impact}.
This paper uses the entire data set and re-analyses it through the lens of our research questions, with an agile vehicle-level feedback loop beyond individual teams as a pervading concern.
To analyze the data, two of the authors first read through all the transcripts and took notes of topics in scope of our lens, to get an overview of the responses and become familiar with the data.
The interviews were then purposefully coded, where respondents covered any of these topics.
The coding scheme was refined in iterations: after coding a number of interviews, the resulting themes were discussed among the authors to identify unclear, overlapping, or missing concepts.
The already coded interviews were then re-coded, and additional interviews then coded by the updated scheme.
Once the resulting coding scheme had been applied to all interviews, the authors together elaborated on the description of the themes, and picked exemplar quotes.
We then aggregated the codes into overall answers to each research question.

The quotes used to exemplify themes have been translated to English, with runs of spoken language edited for readability.
However, to reflect the informal tone of the interviews, colloquial expressions have been kept.

\subsection{Validity and Limitations}\label{sec:limits}
We structure our reasoning about the validity and limitations of the study in accordance with the scheme proposed by Runeson and Höst~\cite{runeson_case_study}.

\subsubsection{Internal validity}

To avoid leading questions, we spent time discussing the phrasing of the questions and types of questions to avoid altogether.
Furthermore, the interview guide was refined through multiple iterations, with input from senior industry experts.
Still, despite making an effort to ask questions neutrally, respondents might have considered mainly negative aspects of their current ways of working.
With agile transformations underway, this could give a positive inclination towards agile as forward-looking and implicitly better.
To provide an in-depth view, we, therefore, both look at how respondents propose that a feedback loop beyond individual teams can be established (RQ2), and, crucially, which implications they foresee from this (RQ3).
Furthermore, we discuss our interpretation of the results and their implications (Section~\ref{sec:interpretation}).

\subsubsection{External validity}
Statistical generalization to a population is not the purpose of a case study.
Rather, any generalization would be \emph{analytical}; that is, for other case with some characteristics comparable to our described one, the results may generalize, but further studies are required for validation.
To enable comparison with other cases, we detail above our case, its context, and our respondent selection.
Furthermore, all interviews were done within one country with automotive companies.
The software developed is largely embedded and expected to have a long lifetime.
Therefore, our findings may not apply to smaller companies, other countries, or for software with a shorter life expectancy. 
Both case companies have global presence, but cultural aspects may persist and could have an impact on how practitioners reason about ways-of-working.
Moreover, we focused the study on one system area at each company, selected for being the most software development intensive ones at the respective companies.
Although we complemented with additional respondents from other system areas, an in-depth study of another area could uncover further detail, possibly contradicting our findings.

\subsubsection{Construct validity}
All authors have prior experience with the automotive domain, having over the last ten years participated in several different automotive research projects, both in Europe and North America.
With both case companies we also have longstanding collaborations, preceding the research project in which this study was conducted.
We leveraged this background knowledge when constructing the interview guide.
Previous experience informed the choice of questions and topics for a first version, which was then refined through multiple iterations, with input from senior industry experts.

Additionally, at the time the study data were collected, the co-interviewers were working at the case companies.
The interview situations were thus informal and characterized by mutual trust.

\subsubsection{Reliability}
To increase reliability, we used observer triangulation during the interviews.
The first author conducted the interviews, joined by one or more of the industrial co-interviewers.
The co-interviewers observed and asked follow-up questions for additional clarifications, to ensure a shared understanding of both questions and answers between respondent and researchers.

\begin{figure*}
    \centering
    \includegraphics[width=0.8\textwidth]{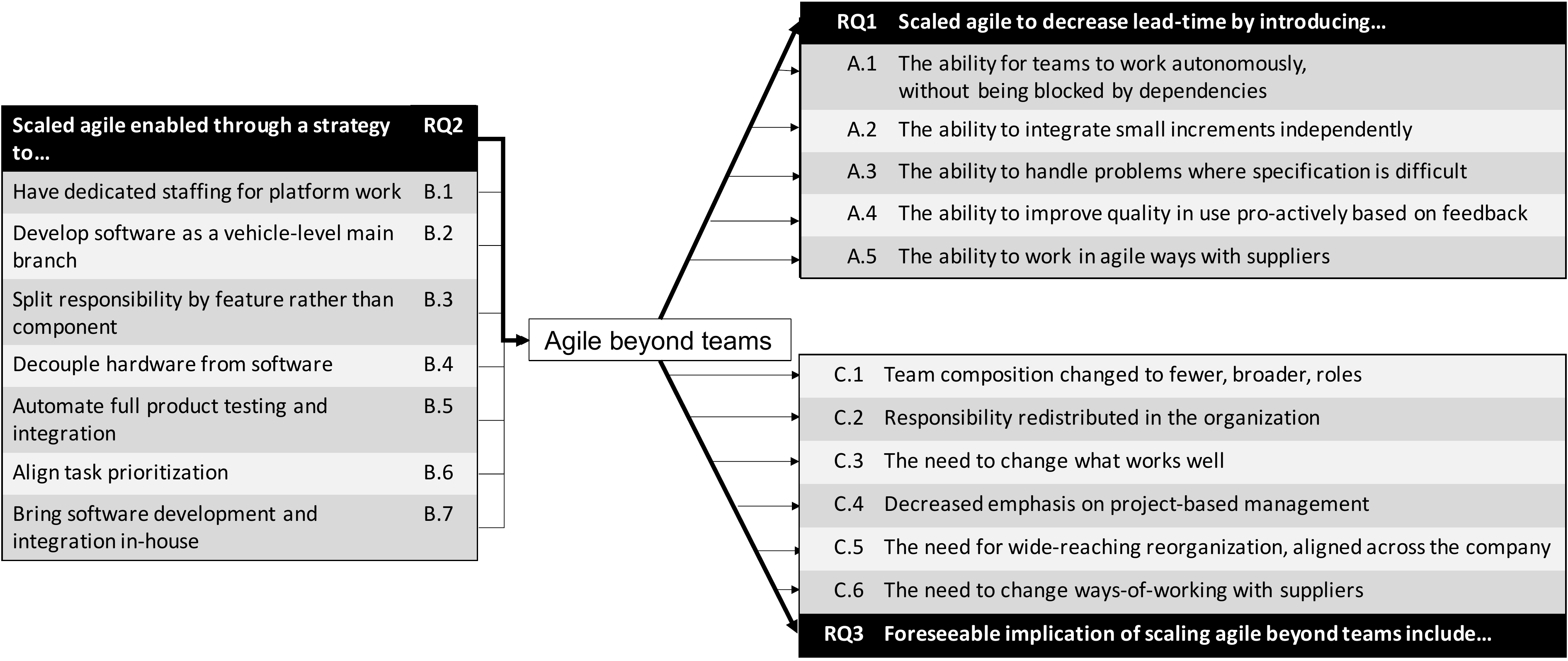}
    \caption{Themes with respect to the three research questions}
    \label{fig:themes}
\end{figure*}

\section{Results}\label{sec:results}

For each research question, we describe the themes that emerged from applying our research method. 
For selected themes, we provide example quotes from our interviews to illustrate the theme with the respondents' own words and to demonstrate our analysis process. %
Figure~\ref{fig:themes} provides an overview:
We identify a number of abilities our case companies are seeking to achieve through an agile vehicle-level feedback loop beyond individual teams (RQ1, top right).
To enable these abilities, respondents propose a number of strategies (RQ2, left); from which they also foresee a number of implications (RQ3, bottom right).

\subsection{Abilities sought through an agile vehicle-level feedback loop beyond individual teams}%
\begin{tcolorbox}[float, title=RQ1 Summary]
Automotive OEMs seek to achieve the following abilities through an agile vehicle-level feedback beyond individual teams:
\begin{itemize}
\item The ability for teams to work autonomously, without being blocked by dependencies
\item The ability to integrate small increments independently
\item The ability to handle problems where specification is difficult
\item The ability to improve quality in use pro-actively based on feedback
\item The ability to work in agile ways with suppliers
\end{itemize}
\end{tcolorbox}

We find several abilities that our case companies seek to achieve through an agile vehicle-level feedback beyond individual teams.
The identified abilities reflect our respondents' views on how scaled agility will help to decrease lead time -- the time until a function can be released to customers. 

\begin{quote}
``In principle you could say that there are a few driving forces for why we need to do things faster than what our current way-of-working allows.
In some areas, the development is very much faster than what a traditional car project is.
Infotainment is a good area; I usually take the example of our speech functionality.
Currently, we specify many years in advance; five years ago, when we specified the digital assistants working today, there where almost no existing digital assistants.'' -- R11
\end{quote}

There is a need to handle increasingly complex vehicles and an increasing share of the customer value being created in software-intensive parts. %
However, respondents do not universally consider the extent of the overall goal agreed on.
They indicate that sentiment in the companies range from that it suffices to make smaller adjustments to the organizations, to that larger reorganizations are necessary. %
Nevertheless, efficient ways-of-working is a shared goal. %
In the following, we elaborate further on the outlined abilities.

\subsubsection{The ability for teams to work autonomously, without being blocked by dependencies}
Agility suggests increased responsibility for teams, allowing teams to work autonomously and independently develop their part of the overall functionality.
The ability to rely on autonomous teams is intended to improve integration, remove negative impacts of technical dependencies, and avoid that problems discovered during integration lead to broadly distributed rework.

\begin{quote}
``The hope is that each team can develop their features [individually]. Ensuring that when each [Simulink] model works, then the combination works.
Before, each team was doing its own model and then testing that it all worked together. And it didn't and we had to redo.'' -- R15
\end{quote}

Thus, it is not enough to aim for autonomous teams, but also the organization must gain an ability to integrate increments from agile teams.
From the perspective of our interviewees, the foundation for this ability needs to be laid within architecture, product structure, and appropriate standardization.  %
In addition, teams and individual developers need to gain good product knowledge, since without it, they will not be able to make good autonomous decisions. %

\begin{tcolorbox}[breakable,enhanced,colback=white!2!white,colframe=gray!35!white]
{\bf In the Literature}:
{Informal risk management is rated as good enough for one autonomous team, while more formal approaches are needed, when several teams work on the same requirement~\cite{Radtke2020}. When several teams work in parallel, Sch{\"o}n et al. conclude that it is beneficial to adopt hybrid development approaches consisting of agile practices and traditional practices~\cite{Radtke2020}.
On the contrary, we observed that our companies are trying to move away from hybrid development approaches. Instead, they rely on architecture, product structure, and appropriate standardization as solutions for supporting antonomous teams to work independently, but at the same time enable integration of increments from agile teams.
} 
{Similar to our findings, Mikalsen et al. argue that distribution of operational tasks with many dependencies as well as misaligned control structures negatively impact agile team autonomy and development speed~\cite{Mikalsen2019}.}
{In general, team autonomy has been identified as key in achieving agility~\cite{Lee2010}.}
However, Paasivaara et al. note that, ``giving teams autonomy without enough coaching led to a suboptimal agile implementation in the teams''~\cite{Paasivaara2018}. 
The second international workshop at XP2019 on autonomous teams aimed at investigating barriers for team autonomy and the main identified barriers for autonomous teams are (ordered list): (i) too many dependencies on others, (ii) lack of trust, and (iii) part-time resources~\cite{Moe2019}. %
{Sekitoleko et al. describe in particular the challenges that can arise in scaled-agile from such technical dependencies and suggest a tradeoff between upfront planning and just-in-time resolutions of dependencies~\cite{Sekitoleko2014}.}
{We conclude that team autonomy indeed promises to positively affect lead-time, if sufficient training is provided and technical dependencies are managed strategically.}
\end{tcolorbox}

\subsubsection{The ability to integrate small increments independently}\label{sec:smallIncrements} %

Respondents report that the feedback cycle time for the complete vehicle system can range up to 20 weeks. %
Shortening feedback cycle time is hence an aspect of shortening the overall development lead-time.
The intention is for teams to receive feedback from integration at the level of the entire vehicle.
Isolated agile teams can already perform continuous integration within their local development, but only to the level of the component they deliver.
The feedback is thus limited to what testing (and test automation) can cover within the scope of that team.
It is important to not only have good flow within individual teams, but also a good rhythm across inter-dependent teams throughout the automotive value chain.
Avoiding negative effects of organizational silos -- business divisions that operate independently and avoid sharing information -- is therefore crucial.

\begin{quote}
``According to the process, each silo is responsible for time, technology, cost. You are responsible for your schedule, your technology, and your cost, which means that you sub-optimize for that. And no one has the job to check that the whole is optimal. Such sub-optimizations inevitably lead to taking these shortcuts that we were talking about, which in turn slows down the speed on the whole.'' -- R7
\end{quote}

This can lead to an undesirable situation, where all teams are fully utilized and showing good progress individually, but not necessarily in a way that permits full features to be integrated at the vehicle level.

\begin{quote}
``What will happen then? Well, then it turns out that if many have made 40\% of their delivery, but no one has made the same 40\%, then we've got around nothing.'' -- R2
\end{quote}

For this reason, an important goal in the OEMs' agile transformation is to take into account the collaboration between teams.
Not utilizing each team to the maximum, but optimize on the whole development.
This requires both organisational and architectural changes, a challenging task, given the historical way of working and the diverse teams spread over organizational silos.

\begin{quote}
``If you are responsible for a function which is spread over 18 nodes, do you want to wait until once half a year to know whether it works? No, exactly. You don't want that.'' -- R4
\end{quote}

Our interviewees express the hope that the ability to have a good flow of independent deliveries will contribute to significantly reducing lead time.
While this ability is generally a good match for an agile transformation, the automotive systems domain demands for additional considerations.
For example, it will be critical to avoid task switching and to provide fast feedback to developers at a time when they still remember the work item in question from the full product.
This will allow to develop less software that does not make it into the car because it does not integrate well.

\begin{tcolorbox}[breakable,enhanced,colback=white!2!white,colframe=gray!35!white]
{\bf In the Literature}: Integration of increments is recognized to be a major challenge in parallel development in multiple teams~\cite{8792665}. Our study also highlights the importance and the challenge of integrating increments into the full product. 
The work in~\cite{STAHL2017150} concludes that direct integration with the mainline increases continuity, but larger organizations are unable or unwilling to work in such a way. Our companies shown a clear interest in minimizing the time from when a feature is conceived to when it is integrated to the full product, even though it is challenging.

The work in~\cite{martensson2016} highlights the risk of the establishment of silo behaviours, when teams tend to establish their own ways of working and  to  see  their software as their own system while treating the complete system as a secondary concern. This is in line with our findings.
In our study we observed the importance of architecture to facilitate integration. This is close to observations in~\cite{martensson2016}, where the authors put the focus on architectural runway, defined in~\cite{1963397} as the infrastructure allowing incorporation of new requirements (new functionalities). The authors also state that ``end-to-end testing is impossible without architectural runway"~\cite{martensson2016}.
\end{tcolorbox}

\subsubsection{The ability to handle problems where specification is difficult}
In contrast to agile ways-of-working, stage-gate development processes emphasize the importance of precise specifications, typically created before development.
For open-ended problems, where up-front specification is difficult, an agile approach to explore the problem space is seen as promising.
This aspect is exacerbated in complex systems requiring large teams working on it.
In stage-gate development processes, a typical approach consists in decomposing a system into subsystems and/or modules, which are then assigned to the various teams.
When working under uncertainty, where the up-front specification is difficult, this way of working is ineffective.
Instead, agile ways-of-working encompassing more than single teams seem to be more appropriate because of their flexibility. %
\begin{quote}
``Very often, when we do our lessons learnt [...] we conclude that we should have started earlier and worked more with the specification. 
But instead of three years, should we bring a four year old product to the market?
I don't think that is the solution.'' -- R11
\end{quote}

While there might be some aspects of system engineering, where long turn-around time is acceptable, there are also those where 20 months and more are problematic.
This includes the development of infotainment components, that must interact with modern phones and will be compared with their modern interfaces, but also the very quickly evolving field of advanced driver support systems.

An additional aspect is that it can be very difficult to anticipate how a high-level requirement can be broken down into system and component requirements.
\begin{quote}
``Much of the development phase is about understanding the task and what the function is, and then bring out some type of technical solution. And it's very much about understanding if the technical solution actually worked as intended.'' -- R5
\end{quote}

In such cases,  a more exploratory approach must be chosen, and a scaled-agile setup promises to provide a better environment due to its ability to iterate over problem-understanding and solution-providing.
This will ideally allow including experience with the technical solution from the street in its design and evolution, thus increasing its fit-for-purpose.

\begin{tcolorbox}[breakable,enhanced,colback=white!2!white,colframe=gray!35!white]
{\bf In the Literature}:
For what concerns dealing with uncertainty, stage-gate and agile development processes build on opposing principles in terms of how they advocate managing uncertainty in new product development. As highlighted in~\cite{BIANCHI2020538}, on one side mixing the two approaches can generate fundamental inconsistencies, on the other side they can coexist since they act at different levels: stage-sate acts as a macro-level framework facilitating the coordination of different teams, whereas agile acts at the micro-level offering effective planning of day-to-day activities~\cite{Runeson06}. 

It is interesting to connect the concept of complexity with the concept of irreversibility.
As discussed by Fowler \cite{Fowler2003}, irreversibility has been identified as one of the prime drivers of complexity.
Agile methods permit to contain complexity by reducing irreversibility. This adds an interesting perspective to our finding on agility being a nice instrument for handling  problems  where  specification  is difficult.
Scaled agile frameworks such as SAFe and LeSS promote practices such as set-based design~\cite[p.75,190]{knaster2017safe} and avoiding narrow product definitions~\cite[p.159]{larman2016large}.
A promising approach to scale benefits of agile approaches beyond teams is to refer to the concept of boundary objects~\cite{Sedano2019,wohlrab2019boundary}, i.e. objects that are relevant for more than one team, such as a product backlog or an agreed-on list of high-level requirements.
Besides of the benefits coming from the flexibility and the iterative nature of agile ways-of-working, boundary objects  would allow for aligning across organizational boundaries as the understanding of problem-to-solve and appropriate technical solutions grows.

\end{tcolorbox}

\subsubsection{The ability to improve quality-in-use pro-actively based on feedback}\label{sec:quality} %
Our respondents strive to achieve a responsive, proactive approach to quality.
This is informed by ideas also present in continuous ways-of-working, but not necessarily the same.
Quality is of course extensively worked on already in the automotive domain.
This theme concerns the new aspect of a proactive, feedback-driven approach, to uphold quality while having short lead-times.

\begin{quote}
``%
There's a good intention in the process, but it gets too slow to work in reality, and then we start taking shortcuts in a more uncontrolled way.
I think the big strength of CI is that you get out of the dilemma of being fast and still having good quality.'' -- R11
\end{quote}

The intention is thus to be able to bring new features to market quickly, without compromising quality.
Quality here also includes achieve quality-in-use -- that customers perceive features as relevant and useful.
Aiming for continuous integration on the entire product as a way to achieve this, it is necessary to scale it beyond single teams.

\begin{tcolorbox}[breakable,enhanced,colback=white!2!white,colframe=gray!35!white]
{\bf In the Literature}:
Some works show that there is 
a reduction in the number of defects when agile is used~\cite{1333387,1359793,1611951}. Another study~\cite{1232877} does 
not find significant differences in either internal or external quality between agile and waterfall models.
We find no reasons to contradict the expectations of our interviewees that, for cases where the full product is hard to specify and its usage is hard to anticipate, iterative feedback from the field and evolution can be superior to upfront specification.
The work in~\cite{martensson2017} proposes a method to integrate automated testing with exploratory testing, i.e. instead of testing with a clear idea of the outcome of the testing, observing the behaviour of the system and evaluating it according with the testers expertise and experience. The authors believe that the two testing approach complement each other and mitigate their weaknesses. This can be a good hint for enhancing the quality of the system. The work in~\cite{martensson2017} also recommend to use scenario-based testing in order to involve end-users. This is in line with our finding of taking into account also the quality-in-use that should be evaluated by end-users.
\end{tcolorbox}

\subsubsection{The ability to work in agile ways with suppliers}

In a complex domain as automotive, there is an incentive to work with suppliers, since it frees the OEM from developing all the software.
However, it is not without cost in terms of lead times and misunderstandings, since procurement is often based on long and complex textual contracts.
In addition, the OEM become dependent on the solution design from the supplier, leading to less control of the software for the OEM. 

Our respondents express that,  seeking agile ways-of-working with suppliers, the aim is to shift focus from contracts to collaboration.
This mainly concerns software where innovation is happening and it is crucial for the OEM to take part in forming the software.
Standardized software -- COTS (Commercial off-the-shelf) -- for functions not regarded as core business, is something our respondents propose buying from suppliers.

\begin{quote}
``So, a Bluetooth stack... I would think that it is the type of software that can be bought. Then there are parts where we need to have quite a lot of collaboration, where we need to be involved and develop the software. We may not have to do everything ourselves, but we need to be involved and give our understanding to the software team outside [the OEM]. Then there is a third type of software and these are the parts you need to make yourself at home.'' -- R4
\end{quote}

Accordingly, a more agile way-of-working between OEM and suppliers is sought for improving the ability to innovate.
But also with the intention to reduce lead-times and thereby improve quality of work artifacts early on in the development cycle.

\begin{tcolorbox}[breakable,enhanced,colback=white!2!white,colframe=gray!35!white]
{\bf In the Literature}:
The work in~\cite{Hohl2016} identifies among the key challenges in agile adoption the collaboration with suppliers. 
The work in~\cite{TransparencyAndContracts} investigates challenges and impediments in continuous software development that involves software suppliers besides of an OEM. Our findings confirm what reported in
\cite{TransparencyAndContracts}: legal contracts are an impediment when scaling agile beyond of the company boundaries.
This study also finds that inter-organizational transparency is considered as positive, especially for what concerns information sharing among different companies, even though not a necessary condition.
\end{tcolorbox}

\subsection{Establishing an agile vehicle-level feedback loop beyond individual teams}

\begin{tcolorbox}[float, title=RQ2 Summary]
Automotive OEMs propose the following for establishing an agile vehicle-level feedback beyond individual teams:
\begin{itemize}
\item Have dedicated staffing for platform work
\item Develop software as a vehicle-level main branch
\item Split responsibility by feature rather than component
\item Decouple hardware from software
\item Automate full product testing and integration
\item Align task prioritization
\item Bring software development and integration in-house
\end{itemize}
\end{tcolorbox}

Our respondents' proposals for establishing an agile vehicle-level feedback beyond individual teams concern both technical and organizational aspects.

\subsubsection{Have dedicated staffing for platform work}\label{sec:rq2:platform}
Automotive software development is typically organized around development for specific vehicle models, in project with fixed scope and a clear end time.
This differs from the agile aim of potentially indefinite continuing work, expressed in the manifesto as ``\emph{The sponsors, developers, and users should be able to maintain a constant pace indefinitely.}"
Handling resources, both human and financial, mainly by projects for specific vehicle models is regarded as problematic by our respondents. %
It is considered to lead to a lack of attention to the shared parts.
Respondents describe this approach to software development as based on how mechanical parts are developed, which is regarded as a poor fit for maintaining software over time.

\begin{quote}
``No-one gets money to continuously maintain their software and then release it to a [specific] project. No, you get money from the project to code a function, as if your software was only that code, not existing anywhere else. And that's how it is in the mechanical world. You get money to change your tool, if we're to have a new trunk lid. If the trunk lid is unchanged you get now money.'' -- R16
\end{quote}

Respondents also indicate that the number of simultaneous projects can be quite high, leading to a loss of overview from the developers.

\begin{quote}
``We can have perhaps 50 projects working against the same product. And the engineers who are to do the change have many projects to consider. It's still the same engineers and the same product that is to be changed. That's a lot of overhead. I would rather see that you develop the product and have ownership and can work more autonomously, no doubt.'' -- R12
\end{quote}

They instead suggest developing a common core as a product, with product ownership of this over time.
The common core would then on request deliver to specific vehicle variants. %
Accordingly, the suggestion to have dedicated staff for platform development is intended to scale agile beyond individual teams.
In particular, it supports the agile value of working continuously at a consistent pace at scale.

\begin{tcolorbox}[breakable,enhanced,colback=white!2!white,colframe=gray!35!white]
{\bf In the Literature}:
The work in~\cite{Klunder2019} surveys the state of the art in the transformation of large companies towards agile software product line engineering. The study highlights that establishing an agile development process that preserves the benefits resulting from already existing SPLs is not trivial. One of the main recommendations of the work is to maintain domain and application engineering during the transformation. This is in line with our finding: our interviewees aim at developing a common core with ownership over time and then to deliver to specific vehicle variants.
\end{tcolorbox}

\subsubsection{Develop software as a vehicle-level main branch}

As per the previous theme, dedicated staffing of platform work is one suggestion from our respondents.
They also point out, however, that such a platform itself needs to be stable, for continuous delivery to development of specific vehicles to be possible.

\begin{quote}
    ``If you are to take continuous integration all the way to a car platform and be able to talk about a backlog and prioritize that, then a prerequisite is that you have a stable base, that we always have a functioning car to work with, and add increments to that functioning car.'' -- R6
\end{quote}

In addition to stability, respondents mention that the platform also needs to allow working with small increments.
Current platforms are not perceived to support this well.
As a mitigation, respondents propose that development and integration should be done as a main track.
Projects for specific car models would then be based on the latest stable release from the main track.

\begin{quote}
``Moving from the project-based development of today to a more product-line-oriented development.
Today, each car [model] project in principle builds their own electrical system, and the similarity between different cars can be big or small, but there are almost always some differences.
We need to get away from that.
We need to have a main track,  and develop that over time so there's always a maturity that makes the entirety fit together.
[For a car project] you then deliver that which is mature once the project is finished.'' -- R7
\end{quote}

Respondents also propose that the development and integration should target interfaces, rather than specific components.
If the vehicle is defined in terms of components, the view is that problems in the interaction between components are discovered only late in integration.
Focusing on interfaces rather than component content is also viewed as a way to make the dependencies between teams explicit.
This relates also to the strategy of feature-oriented, rather than component-oriented development.

\begin{tcolorbox}[breakable,enhanced,colback=white!2!white,colframe=gray!35!white]
{\bf In the Literature}:
This discussion is related to the one in Section~\ref{sec:smallIncrements}. The work in~\cite{STAHL2017150} investigates the correlation between the size of software development efforts and the ability to practice continuous integration. These correlations are: (i) software size negatively affects the continuity of continuous integration, (ii) there is a clear negative correlation between organizational size and continuity, (iii) continuity correlates positively with the proportion of developers in the organization, (iv) breaking down large systems into smaller pieces is a key enabler for continuous integration at scale, and (v) direct integration with the mainline increases continuity but larger organizations are unable or unwilling to work in such a way. Contrariwise, our companies aim towards the development of software as a system-level main branch and direct integration with it. The work in~\cite{MARTENSSON2018223} identifies twelve factors that affect how often developers deliver software to the mainline. These twelve factors and grouped in four themes: activity planning and execution, system thinking, speed, and confidence through test activities. We found various similarities with these findings in our study. We also found a system thinking recommendation, in terms of producing a modular and loosely coupled architecture and developers must think about the complete system and interfaces instead than to components content. In terms of speed, we also found that fast feedback from the integration pipeline is probably the main motivator to commit frequently.
\end{tcolorbox}

\subsubsection{Split responsibility by feature rather than component}
Both case companies currently have vehicle architectures structured mainly by hardware components -- Electronic Control Units (ECUs).
The division of responsibility in the organizations also largely follows this structure.
Functions where the implementation spans multiple ECUs can thus involve many teams, with each team implementing the part of the software that will run on the ECU they are responsible for.
Our respondents suggest that instead dividing responsibility by features, and having teams be responsible end-to-end for its integration in the vehicle, could increase team autonomy, and thereby the overall agility in the organization.
This way of dividing responsibility is sometimes termed vertical slicing.
Another aspect raised by respondents is that the responsibility of the team should be stable over time, so that the team can take ownership of their task, and avoid frequent task switching.
To facilitate architecting for a division by features, respondents also suggest striving for Commercial Off-The-Shelf (COTS) components for non-OEM specific functionality.

\begin{tcolorbox}[breakable,enhanced,colback=white!2!white,colframe=gray!35!white]
{\bf In the Literature}:
Feature and component teams have been discussed in the context of scaled agile \cite{leffingwell2010agile}, and, if the system under construction allows for it, feature teams are generally recommended.
Kasauli et al. compare feature and component teams with respect to requirements challenges \cite{Kasauli2020b}. 
They find that, in large systems companies, deep knowledge on some components may be required and could be maintained in a component team, but may require a more traditional requirements breakdown. 
In contrast, feature teams may be more suitable to overcome silos and to tackle problems that are hard to specify, but also be more prone to introduce technical debt and to lack a system or component perspective.
Our findings are related to the findings in~\cite{MARTENSSON2018223}. In particular, the work highlights that the organization is more important than the architecture. When explaining that the project organization must both work with small changes and take responsibility for the architecture, the authors explain that the problem in not in the components, but in the fact of having component teams.
\end{tcolorbox}

\subsubsection{Decouple hardware from software}
Manufacturing of hardware (electronics) and mechanical parts have much longer lead time than software.
Physical prototypes are also costly to produce, whereas software inherently remains virtual, also once deployed.
Respondents suggest decoupling hardware and software development as a way to separate tasks along how straightforwardly agile ways-of-working can be applied.
By separating software from specific hardware, rather than having it bundled and delivered as part of hardware components, software is untied from the inherently longer hardware development loop.
Vehicle-level software integration can then happen during sprints, as software delivery does not involve hardware delivery.

\begin{quote}
``We have a quite tight coupling between hardware and software. Even to the point that we release software as a part of hardware. But that also has us waiting for hardware, to verify that the software works. We need to work our way out of that.'' -- R11
\end{quote}

A larger part of the vehicle software development then resembles development for pure software systems.%

\begin{tcolorbox}[breakable,enhanced,colback=white!2!white,colframe=gray!35!white]
{\bf In the Literature}:
The work in~\cite{9213824} highligts that decoupling software from hardware improves the system upgradeability and scalability. The work in~\cite{7930217}
reports on a survey with system architects, which aims at understanding to what extent SOA concepts are applicable for safety-critical embedded automotive software systems. 
They recommend to represent vehicle functions as services and services should be designed independently from a target hardware platform.
\end{tcolorbox}

\subsubsection{Automate full product testing and integration}
Automatic testing is a staple of agile ways-of-working.
Increasing the test automation is a recurring proposal from our respondents, particularly on levels involving multiple teams.
While teams can currently continuously integrate software towards the component they cover, continuous integration to the level of the entire vehicle is currently a considerably longer loop.
This integration happens in multiple levels, from component, via subsystems, to the whole vehicle.
Given the complexity of automotive systems, respondents regard the multiple levels as a necessity, but where the levels outside single teams need a considerably higher degree of automated test than what is currently available.

\begin{quote}
  ``Getting to a degree of automation that lets you develop your little part and bring that into the [vehicle] and get a response, preferably automated, back: `What should I do to make it work?' We don't quite do that today. [Development] is decomposed all the way to micro components and then verified all the way back up the other leg [of the process]. If you include software and hardware, then it's a rather long development and verification loop.'' -- R5
\end{quote}

\begin{tcolorbox}[breakable,enhanced,colback=white!2!white,colframe=gray!35!white]
{\bf In the Literature}:
The work in~\cite{Klunder2019} relates the testing strategy with a software product line approach and highlights the need for an adjustable and scalable test strategy that enables short iterations and for all variants. Moreover, the same work highlights the  need for a virtual integration.
The work in~\cite{9112168} discusses about cross-ECU testing required by the fact that there are dependencies among different ECUs. In particular, this work highlights that cross-ECU testing might be challenging because of not synchronised development process of the different ECUs. 
The work in~\cite{martensson2017} highlights the benefit of integrating automate testing with (manual) exploratory testing. The work in~\cite{MARTENSSON2021110890} testifies also that exploratory testing is a good instrument to test system-wide and to test large-scale systems, especially when performed with an end-user perspective. The fact of having and end-user perspective is highlighted also in Section~\ref{sec:quality} as an important aspect for quality, and specifically for quality-in-use.
\end{tcolorbox}

\subsubsection{Align task prioritization}
Agile methods emphasize delivering value continuously in small increments.
To enable incremental deliveries to span multiple teams, overall goals and priorities need to be aligned between teams.
Furthermore, respondents point to the importance of coordinating and aligning on a high abstraction level.
One way to coordinate between teams, without sacrificing agile methods within the teams, is through shared task prioritization in centralized backlogs.
However, respondents point out that aiming for systematic structuring can take uniformity to where it becomes counterproductive.
It is important to still give the teams the freedom to define their ways-of-working and allow local variation where alignment is not needed.

\begin{quote}
``Sometimes I think we want to do everything so uniform and that I think can be counter-productive. /…/ Rather that [Continuous Integration] can grow bottom-up and that you meet up there some way. Instead of pushing it top-down.'' -- R18
\end{quote}

\begin{tcolorbox}[breakable,enhanced,colback=white!2!white,colframe=gray!35!white]
{\bf In the Literature}:
One of the strength of SAFe\footnote{\url{https://www.scaledagileframework.com/}} is in providing a clear breakdown hierarchy from enterprise level to teams \cite{knaster2017safe}. When combined with awareness on how each part fits in the complete picture, this might discourage localized thinking. As testified by our study, this could also have bad effects. Our respondents, in fact, highlight that it is important to give freedom to the teams to define their ways-of-working. 
\end{tcolorbox}

\subsubsection{Bring software development and integration in-house}
Respondents mention bringing software development in-house as a way to achieve a faster development loop.
While it is regarded as neither desirable nor realistic for an OEM to develop all the software in a vehicle, respondents also raise the need to be strategic about which software functions an OEM should develop in-house.
Vehicle-level control functions are in particular brought up as one such type of software.
Respondents also prefer having integration in-house; buying the integration of software (for example for an ECU) is seen as leading to slower feedback.

\begin{quote}
``There is a lot to do, if we only talk software, but everything cannot be living all the time.
That which is living all the time we should do as close as possible, whether we do it ourselves or have a joint venture.
But there the communication distance needs to be short, so it's possible to iterate fast.'' -- R3
\end{quote}

\begin{tcolorbox}[breakable,enhanced,colback=white!2!white,colframe=gray!35!white]
{\bf In the Literature}:
The work in~\cite{Hohl2016} advises more in-house software development as a solution to the challenge of working with suppliers when transitioning towards agile software development.
The importance to shift towards more in-house software development is highlighted also in~\cite{aagren2019impact}, where the authors identify in increased flexibility and ability to quickly relate to changes the most important benefits of this shift.
The work in~\cite{ICSA2021} highlights, among the main business goals of automotive OEMs, the opening of their platforms to third-party companies with little to no knowledge of automotive systems. This would lead to the creation of a software ecosystem, similar to what we can observe in the smartphones domain. However, at the same  time, automotive  OEMs  aim  at increasing  control  over  concerns  and  unknowns. This is considered as a  crucial step for their innovation capabilities. These two contrasting forces might lead to the outsourcing of more standardised components, while keeping under the control of the company the components more innovative and that are more uncertain. 
\end{tcolorbox}

\subsection{Implications of the proposed ways of introducing an agile vehicle-level feedback loop beyond individual teams}
For the proposed strategies, our respondents also elaborate on foreseen implications. %
The identified implications all come with opportunities, risks, and trade-offs.

\begin{tcolorbox}[float, title=RQ3 Summary]
Automotive OEMs foresee the following implications from the proposed ways of introducing an agile vehicle-level feedback loop beyond individual teams:
\begin{itemize}
\item Team composition changed to fewer, broader, roles
\item Responsibility redistributed in the organization
\item The need to change what works well
\item Decreased emphasis on project-based management %
\item The need for wide-reaching reorganization, aligned across the company
\item The need to change ways-of-working with suppliers
\end{itemize}
\end{tcolorbox}

\subsubsection{Team composition changed to fewer, broader, roles}
Respondents point out that the ability for teams to work autonomously calls for teams of generalists.
Bringing vertical slices -- features of end-to-end functionality -- from development to integration in the vehicle involves many competencies, and team members thus together need to cover a wide range of skills.
Fitting all needed competencies in one team can be practically infeasible. %
However, strictly area-of-competence-based specialist teams are not regarded by respondents as autonomous in the agile sense.
\begin{quote}
``If you put experts from one specific area of competence in one group, and other experts in another group to be strictly competence-oriented, then you don't have these autonomous groups who can develop their functionality with testing, software development, and algorithm competencies gathered in one group.'' -- R5
\end{quote}

Creating autonomous teams in automotive requires a multitude of different competencies.
The suggested team setup has implications on architecture, both to enable test automation to product integration -- the vehicle-level, and to allow teams to work on features, rather than specific hardware components.

Access to experts could become a bottleneck; teams organization itself will not change the available amount of experts in the organization.
In relation to this, respondents also bring up a possible tension between a team intended to be autonomous, and external experts provided by the surrounding organization. 
Corrections coming from outside the team risk being regarded as infringing on the team's mandate.

\begin{quote}
``If you want an autonomous team, then they want the mandate to develop their own product, and if you don't have included the right competency from the start, conflicts arise when someone comes from the outside an says `But that thing, that wasn't good'.'' -- R13
\end{quote}

With the current, highly-specialized roles, a single person can have multiple roles, to cover the entirety of the expertise they provide to the organization.
Re-organizing to have persons tied to teams can clash with this, placing only part of a persons roles inside the team.

A reduction of the number of roles is, however, brought up as positive by our respondents.
The sheer amount of current roles is regarded as problematic in itself.
While the ideas behind the existing roles are seen as sound, in practice creating all of them could be less beneficial.
Handovers between many, narrow, roles, is regarded as a cause of slowness.
Partially because of the amount of stakeholders involved, and partially because current technical solutions are highly interdependent.
This leads to the necessity of involving many roles. %
Respondents, however, caution that vehicle development is a balancing of different, potentially conflicting, interest.
\begin{quote}
``If you just said that `OK, you get a defined function, you have this space on the node, do what you want with it', we would probably have pretty suboptimized cars.'' -- R2
\end{quote}

Our respondents also note that, while agile teams have existed in automotive for a long time, there is now also a top-down initiative for agility on a wider scope.
The feedback loop to a team thus covering a larger extent of the entire development and integration chain than before.
For the roles in the middle layer of this change, for example line managers and project managers, respondents note that the proposed changes may be uncomfortable.
In particular, those who are to carry out the change are those affected the most, sometimes possibly in a negative way, for example if a manager role is to be changed or removed.
This is regarded as particularly challenging to handle.
One respondent suggests, however, that precisely defined roles may be important for the individual to feel confident in their work, and to deliver independently, but for the organization as a whole, roles may be less important.
Additionally, the transparency and visibility that comes with the ability to constantly track delivery status and progress is brought up as uncomfortable for some.

\begin{tcolorbox}[breakable,enhanced,colback=white!2!white,colframe=gray!35!white]
{\bf In the Literature}:
Related to our findings, SAFE~\cite{knaster2017safe} mitigates the difficulty of interaction among people who perform the same role in different and multidisciplinary teams with the introduction of communities of practice (CoPs). It can be seen as a support network for people who share a common role to benefit from each other's experience. According to SAFE, CoPs create opportunities for learning, building capability, sharing knowledge, and, at the same time, reducing risk of duplication of work. However, the work in~\cite{Vestues2016} found  evidence that it is not always obvious that CoPs solves all problems related to knowledge sharing and process improvement. The authors also highlight both the need of dynamic structures in learning and the fact that different approaches can be taken based on the project type. In general, it is not clear how CoPs and similar unofficial arenas for learning and coordination should change over time~\cite{Vestues2016}.

\end{tcolorbox}

\subsubsection{Responsibility redistributed in the organization}
With less focus on detailed requirements and specifications, there is also more ambiguity about responsibility.
Respondents elaborate that, while increased throughput is the intention behind more feedback-driven development, increased ambiguity could be seen as the price.

\begin{quote}
``If you want a more agile process, where you don't document requirements in detail and don't decide the content exactly in advance, then by definition it will be more unclear exactly where responsibility lies. That's in a sense a price you pay for speed.'' -- R11
\end{quote}

When moving from responsibility-following-line-organization to a product owner structure and increased team autonomy, respondents stress the importance that responsibility is indeed moved rather than removed.
They also indicate that during a transition, when a previous organizational structure still lingers, where responsibility follows technology areas and a line organization, it can be difficult to let teams take over the responsibilities.
Another potential challenge brought up is a leadership style of detailed control.
When responsibility resides with a manager role, this can clash with the pursuit of high autonomy for the team.

\begin{quote}
``In the end it's the manager who signs off on the technical delivery, including me, the last one to sign off. This is probably one of the difficulties when we talk about product ownership. Then we're down on leadership. There are leaders who say `if I'm to sign there I need to know everything about everything'. Who command, in principle. That's not a good way to create autonomous teams.'' -- R20
\end{quote}

Introducing a vehicle-level feedback loop does not guarantee that complexities of the vehicle, e.g., dependencies between different parts of the vehicle, will be handled differently than today.
Respondents suggest that some of the complexity on the level of the entire vehicle comes from implicit dependencies between parts. %
They point out that slicing the problem, where to place hand-overs between different areas, is inherently difficult.
Even if subdividing by feature, rather than component, development of a large system will involve coordination between different development groups.

\begin{quote}
``If I'm responsible for a function and there are many different ways to control it, both on and off board. If I'm responsible for the air conditioner in the car should I also be responsible for apps, cloud services, servers\ldots No, that's probably not so good. Then there's a handover somewhere. That's not simple.'' -- R3
\end{quote}
\begin{tcolorbox}[breakable,enhanced,colback=white!2!white,colframe=gray!35!white]
{\bf In the Literature}: Some works in the literature investigated the challenges related to the redistribution of responsibilities in the organization. 
The work in~\cite{Paasivaara2018} reports how Ericsson introduced agile in a new R\&D project while, at the same time, scaling it up. The work highlights challenges in defining the product owner role and in clearly identifying the responsibilities. The work also reports about the role unclarity for middle managers in agile and the resistance of 
management in maintaining the waterfall mode, thus keeping the old bureaucracy and internal silos. 
Th management unwilling to change is one of the challenges highlighted also in the systematic literature review reported in~\cite{DIKERT201687} (found in 4 primary studies). This survey highlights also uncertainty in the definition of middle managers' role (found in 7 primary studies), together with the management sticking in the waterfall mode (found in 6 primary studies), and the keeping of internal silos and bureaucracy (found in two primary studies).
\end{tcolorbox}

\subsubsection{The need to change what works well}
Our respondents are generally positive to adopting agile ways-of-working on a broad company level.
However, they also caution that current, more stage-gated, ways-of-working do turn out quality vehicles, with good results for the companies.

\begin{quote}
``I think that's the difficulty now, that we have a kind of project machine that yet does deliver something all the time.
And we have few manufacturing disturbances. We earn lots of money and the customers get their trucks and are really happy.
We're trying to make that even better, but we don't want to, while doing that, destroy the machinery.'' -- R12
\end{quote}

The current competence-based roles were also once established for reasons of efficiency, similar to the goal for a vehicle-level feedback loop beyond individual teams.
Not derailing something that works is a key concern. %
Respondents also raise the concern that changing the integration structure and thereby the delivery structure may also change power structures, which could trigger resistance. %
Also highlighted is that, while stage-gate processes where software is delivered on few occasions, are at odds with continuous delivery, they may have offered good value on the overall system-level. %
The more software-intensive the vehicles becomes, respondents thus suggest that it becomes necessary to alter what works well, to obtain something better.
However, the proposed kind of feedback loop, as it is setup for example in SAFe, is seen by respondents as needing a stable ground.
In the typical software case, this could be a server infrastructure with a software stack.
Respondents thus question to which extent preexisting development methods that include feedback loops covering an entire system can be applied to automotive as-is.

\begin{tcolorbox}[breakable,enhanced,colback=white!2!white,colframe=gray!35!white]
{\bf In the Literature}: Our respondents are positive towards adopting agile ways-of-working on company level, but at the same time they are aware of the risk of compromising what is working.
This caution is shared with some works we can find in the literature. 
For example, Elbanna and Sarker highlight that agile software development can intensify technical-debt concerns and this accumulation can lead to unexpected project delays and lower software quality~\cite{7325176}.
Our respondents also highlight that there is the need to change what is working to obtain something better.
Indeed that are some challenges in adopting agile ways-of-working on company level, inline with what can be found in the literature.
The work in~\cite{Kalenda2018} identifies resistance to change, quality assurance concerns, and integration into preexisting non-agile business processes among the main critical challenges. 
\end{tcolorbox}

\subsubsection{Decreased emphasis on project-based management}
At the time the data was collected, our case companies mostly organized development as projects, implying a fixed scope and a fixed duration.
In contrast, the envisioned product-based development is neither delimited in scope nor time.
This relates back to theme~\ref{sec:rq2:platform}
Treating what a team delivers as a product they maintain over time decreases the focus on projects when defining deliverables.
Respondents stress that this shift both changes the organization of work and where in the organization power and mandate reside.
Going from a project focus to a product focus impacts when returns on investments can be observed and where prioritization happens.

\begin{quote}
``To develop software as a product demands an investment that doesn't directly reflect what projects order. Because the projects order functions, and that job is quite easy. It's more difficult to maintain that source code over time across projects cost-efficiently.'' -- R15
\end{quote}

Particularly with respect to a portfolio of vehicle models if the majority of development is structured around a shared core.

\begin{quote}
``It's more about the portfolio dimension compared with the specific car. If you develop functionality that's part of all kinds of cars, how can you then see if that job was worth doing or not. And how it then ends up in 15 different car models, where do you understand if ends met or not?'' -- R11
\end{quote}

Additionally, autonomous teams delivering functionality all the way to the vehicle level constitutes a shift in how to work cross-functionally.
\begin{quote}
``It's how we've [previously] set up the organization and the process, we're very governed by projects. The way of working cross-functionally is in projects, you could say'' -- R12
\end{quote}

\begin{tcolorbox}[breakable,enhanced,colback=white!2!white,colframe=gray!35!white]
{\bf In the Literature}:
Kersten discusses how the traditional project culture of many companies is blocking their ability of future successes \cite{Kersten2018}. He claims that many fortune-500 companies will disappear, if they are not able to drastically change their ways-of-working. In particular, he suggests the Flow-Framework to more quickly connect business and technology without the boundaries introduced by cost-centers and project-politics.
\end{tcolorbox}

\subsubsection{The need for wide-reaching reorganization, aligned across the company}
The proposed strategies can have an impact with larger organizational scope than a number of teams, in some cases the entire company.
For example, integration on the full vehicle may require integration practices to be aligned on a detailed technical level between many teams.
The integration can also require intermediate steps, and thus impact several levels of the organization.
This limits how far agile can be scaled bottom-up by initiatives from the teams.

\begin{quote}
``Now you want to do it globally [across the organization]. It doesn't help if one tiny area works in this way, but not the rest.'' -- R5
\end{quote}

Even with support from management, an agile feedback loop including the vehicle-level may require a company-wide initiative.
What starts as software development considerations thus lead to implications on the company business strategy.

\begin{tcolorbox}[breakable,enhanced,colback=white!2!white,colframe=gray!35!white]
{\bf In the Literature}:
SAFE supports the formation of CoPs to align various teams on common needs~\cite{leffingwell2016safe,knaster2017safe}. However, as highlighted in~\cite{KASAULI2021110851}, there is the need of empirical evaluation of these methods in practice.

\end{tcolorbox}

\subsubsection{The need to change ways-of-working with suppliers}
Large parts of vehicle development is currently done by suppliers to the OEMs.
Software is in these cases commonly delivered together with, and integrated in, a hardware part.
A feedback-loop on the vehicle-level thus also involves suppliers, which calls for changed ways of collaborating.
The collaboration between OEM and suppliers is often governed by contracts with detailed specification for deliverables at a few discrete points in time.
Working with more continuous delivery of software would thus affect how compliance is ensured, and have legal implications.
A particularly sensitive question is the division of responsibility in the event of faults or warranty problems.

\begin{quote}
``You try to be as watertight as possible from both ends. There it would probably be better to have more collaboration.
The reason for this type of contractual way-of-working is, who takes responsibility when something goes awry? Who pays the warranty costs and who takes the product responsibility?'' -- R12
\end{quote}

Changing the contractual setup for how software is delivered also affects economic aspects.
Payment is typically based on the number of part purchased, that is, the number of hardware units.
Delivery of software is thus not bought separately.

\begin{quote}
``It's surely in some part contractual, but there's also very much a partnership and relation part. A supplier of software today doesn't earn anything. Developing software is only a cost. What they earn from is selling hardware to us, that's all we pay for.'' -- R3
\end{quote}

\begin{tcolorbox}[breakable,enhanced,colback=white!2!white,colframe=gray!35!white]
{\bf In the Literature}:
The work in~\cite{TransparencyAndContracts} highlights that legal contracts are an impediment when scaling agile beyond of the company boundaries. Contract-based collaboration hinders early feedback from suppliers and can introduce unnecessary costs in the case of changes in the software are needed~\cite{Agren2019}.
\end{tcolorbox}

\section{Discussion}\label{sec:interpretation}
The sought abilities, and the proposed ways of achieving them, draw largely on bringing possibilities from development of software-only systems to the mechatronic systems in automotive.
While this may seem like an unattainable vision because of the unavoidably physical reality of mechanical parts, both hardware and mechanical development are already largely virtualized; for example through the use of CAD (Computer-Aided Design) and crash simulations for mechanical parts, and development of hardware in languages like VHDL and SystemC.
Hypothetically, automotive development could leverage an entirely virtual vehicle, integrating the virtual representations -- models -- from all the development activities.
As noted by our respondents, however, the sheer complexity of automating the tests for the software parts alone is daunting.
Combining models originally created for separate, specific, purposes also comes with its own set of challenges~\cite{agren2019automotive}.
For example, a model created for verifying the timing in a single ECU will then need to maintain external interfaces for use in the full development chain.
A possible approach, in line with the suggestions from our respondents, is then to strive to separate as much as possible those parts of the vehicle development which can be done in software only.
This also suggests aiming for clear abstractions between hardware and software, and, when developing new functions, aiming to do this mainly in software, broadly construed.
While we have not aimed to define agility, our results collate aspects our case companies associate with being agile.
In addition to the themes (Section~\ref{sec:results}) derived directly from the respondents' insights, we also provide our interpretations of concerns cutting across the research questions.
These overarching interpretations aggregate the findings from the themes, per research question, and are thus one step removed from the data.

\subsection{Abilities sought through an agile vehicle-level feedback loop beyond individual teams (RQ1)}
On the company level, decreasing the time to market is a goal that is thought to be supported with transitioning to organization-wide scaled agile.
In the view of our respondents, working in an agile way also at the organizational scope that involves the entire vehicle could mean establishing a short feedback loop, where teams deliver small increments, increments are integrated with the vehicle, and the team receives feedback fast enough to match the typical pace of agile sprints of a few weeks.
The intention is that increments can be both delivered and integrated independently, keeping a consistent flow out from each team.
With frequent integration and feedback on each increment, the hope is to increase team independence, since teams check how each small change works in the context of the full system.
Frequent feedback, in particular from vehicle-level integration, is also seen as way to reduce dependency on specifications.
The ability to validate or refute assumptions close in time to when they were made allows for exploration during development.
Our respondents particularly point to the need for this for problems where upfront specification is difficult.
Exploration and the ability to vet assumptions is also sought in order to ensure that the features developed are relevant to customers.
And, for achieving faster feedback towards suppliers, the aim is also to be able to involve them in these feedback loops.

\subsection{Establishing an agile vehicle-level feedback loop beyond individual teams (RQ2)}
A common thread in respondents' suggestions is to establish a platform or notion of vehicle-level software.
Integration is reported to currently be a lengthy process, and spread out over development.
The suggested vehicle-level software would correspond to a main branch for development;
into which teams would integrate small increments throughout development, rather than providing large deliverables for full vehicle integration only at a few occasions.
The vision for such core development is to bring it in-house, as it is regarded as holding the differentiating features for the OEMs.
In-house development is also seen as faster, with inherently shorter communication distances.
However, respondents point out that a vehicle-level feedback loop could also involve suppliers.
Practically, this could mean suppliers delivering software at a pace of each sprint.
It can also involve co-locating teams, staffed jointly by OEMs and suppliers.
Such new ways of working between OEMs and suppliers do however differs considerably from how current contracts are typically setup.
Respondents point out that this business dimension may overshadow practical and technical concerns.

Respondents also bring up that development of functionality common across vehicle models needs dedicated staffing.
The perceptions is that developing the common vehicle-level only in projects targeting specific car models leads to suboptimizing for the vehicle at hand, rather that for cross-cutting concerns.
The vehicle-level is suggested as decoupled from hardware, to benefit from the shorted development cycles for software.
Where current development is decomposed by physical component -- Electronic Control Unit (ECU) -- the suggestion is here to instead decouple software development from specific hardware.
Instead of defining the vehicle-level in terms of component contents, respondents suggest defining interfaces.

In addition to aiming for decoupling of hardware and software, interfaces instead of component content is intended to allow development to be split by feature, rather than component; in turn aimed at easing integration.
Having a notion of vehicle-level software, and continuously integrating increments with this, is suggested as aiding task prioritization between teams, since work items are defined in terms of a common whole.
Supporting integration at this scale, however, is highlighted as a considerable challenge.
Achieving sufficient test coverage, a sufficient degree of test automation, and handling testing rigs and testing scenarios not originally created for automation, are all technical challenges brought up by respondents.
Alignment of tooling between teams is regarded as necessary to some extent.
Respondents also voice concern, however, that conformity can be taken too far, to where it stifles the desired flexibility for the single teams.

\subsection{Implications of the proposed ways of introducing an agile vehicle-level feedback loop beyond individual teams (RQ3)}

Respondents note that introducing the sought feedback loop may necessitate changes to currently well-performing ways-of-working.
For example, a feedback-driven way-of-working may increase throughput and team autonomy.
However, it can also also increase ambiguity of where in the organization responsibility resides, and decrease the emphasis on precise specification and documentation.
Additionally, the complexity of interdependent parts in the vehicle is not automatically handled any different because a vehicle-level feedback loop is introduced.
A further example is that going from a focus on projects to a focus on sustaining development of a shared platform impacts when returns on investments can be observed and where in the organization prioritization between different possible features happens.

Respondents also anticipate a move to fewer, broader, roles.
A reduction in the number of roles is seen as positive, although a need to fit many competencies in each team is in itself an organizational challenge.

One of the critical challenges within the automotive domain is that strong ownership of code goes against the agile way of developing code.
Code can be spread over many ECUs, each having a different suppliers as responsible.
Changing this may require both contractual and architectural changes.
If a large number of teams are permitted to work on each ECU, there needs to be a precise way to validate the overall quality.
To put most of the platform code into a few ECUs would be another way of supporting a more flexible way-of-working.

The impact of the proposed changes may extend far enough in the organization to become a business strategic decision for the entire company.
Achieving these changes is not mainly technically challenging, but may take time, since the organizations are large and have previously worked plan-driven for many years.
For an extended period, results may worsen before getting better.
However, not doing anything may be an even greater risk, and both our case companies have had agile transformations ongoing for a few years.

\subsection{Summary}
Automotive OEMs are applying agile ways-of-working to the development of the entire vehicle, beyond software development.
Specifically, we investigate the feedback loop of agile development at the vehicle-level; abilities sought, strategies proposed for achieving these, and foreseen implications.
Since this involves more of the organization than software development, the problem becomes multidisciplinary, and what the literature covers for each theme mainly deals with the theme in itself, rather than the intersection with other disciplines.
Figure~\ref{fig:themes} gives an overview of the themes we find for each research question.
To the left, we cover the strategies our respondents propose (RQ2).
These aim to establish the feedback loop by keeping development tightly together at the OEM -- in-house staff with aligned task prioritization developing and integrating a vehicle-platform.
Applying these strategies is intended to yield a number of abilities (RQ1, top right), which however do not come for free; a number of implications are anticipated (RQ3, bottom right).
The implications are not bound exclusively to the proposed strategies, but also follow from the vehicle-level scope of the sought abilities.
As the implications go beyond software development, broadly impacting the organization and possibly also extending to suppliers, an organization needs to carefully consider if anticipated benefits from an agile feedback loop at scale are worth the implications.

In this way, Figure~\ref{fig:themes} is a qualitative model that visualizes the causalities derived from our interview study.

\section{Future Work}\label{sec:future}
In addition to the methodological limitations of the study, which are discussed in Section~\ref{sec:limits}, below we discuss the limitations of our results, and suggest directions future research.
As per our method, only analytical generalizations of the results are possible, and require further studies for validation.
We speculate, however, that our results may hold for other automotive OEMs pursuing agile feedback loops in their vehicle-level development, and potentially for automotive suppliers developing software-intensive mechatronic components.
Companies in other domains with software intensive, but not purely software, systems, for example telecom \cite{Paasivaara2018}, may seek similar abilities applying agile ways-of-working at scale, and thus face similar implications.

For software to not be tied to the development loop of hardware, decoupling the two is necessary.
Our results are, however, limited to indicating that such a split would be beneficial.
More research is needed on how to realize the decoupling.
Some low-level software, for example drivers, will remain closely tied to hardware, and will not be possible to separate from the hardware-development-cycle.
Decoupling might, however, be a prerequisite for supporting platforms with service-oriented software.
If hardware capabilities, for example performance, are limited, this inhibits the possibility of feature growth.
This ties to the research directions of vehicle-centralized architectures (see Section~\ref{sec:future_arch}), aiming to create platforms that support feature growth after the vehicle has been released to market.

While we relate each theme to previous research, variations within the themes could be further investigated.
In particular the strategies suggested by our respondents (RQ2) may vary in applicability, also for comparable cases.
For example, although feature teams are recommended over component teams when possible \cite{leffingwell2010agile}, combinations of the two, when component teams are needed are adaptations to research further.

To achieve the outlined abilities (RQ1), a company may considered the implications foreseen (RQ3) worthwhile.
However, our research is limited to anticipating these implications; how to handle them needs to be explored and validated in future research.
For example, a cross-functional team taking responsibility for features spanning a breadth of the vehicle development will need to cover a breadth of competencies.
Our respondents anticipate that such teams will be the aim, rather than handovers between teams of narrow specialist areas.
How to balance the need for fitting many competencies in a team to achieve team autonomy, with the need for specializations to tackle complex features remains an open research area.

Another important issue are contracts towards suppliers, that are moving more towards trust rather than rigorous contracts \cite{TransparencyAndContracts} hard to understand and fulfil; but altogether removing contracts also causes challenges regarding legal responsibility and disagreement towards when a job is completed. Here more research is needed.

\section{Conclusions}\label{sec:conclusion}
Our respondents are aware that the proposed changes come with risks and will be challenging to achieve.
However, the sought abilities are seen as necessary for handling the increasing amount of software in automotive systems.
The perception is also that the abilities cannot be achieved with previous, plan-driven ways-of-working.

Going beyond single teams in automotive requires going beyond software.
Considerable amounts of software in an automotive system is developed in close connection to hardware development.
From interviews with high- and mid-level managers in two automotive OEMs, we derive qualitative insights on agile ways-of-working beyond individual teams teams.
We analyze what abilities automotive OEMs are seeking to achieve from having the feedback loop of agile methods at this organizational scope;
proposed ways for establishing such as feedback loop, and implications anticipated from the proposed ways.
For managers in automotive or other domains with a mix of software, hardware, and mechatronics, we hope the findings can provide guidance on the trade-offs between  different approaches when scaling agile.

\balance

\bibliographystyle{IEEEtran}

\end{document}